\documentclass[a4paper,twocolumn]{article}
\usepackage[margin=2cm]{geometry}

\usepackage{subcaption}
\usepackage{calc}
\usepackage{amssymb}
\usepackage{amstext}
\usepackage{amsmath}
\usepackage{amsthm}
\usepackage{multicol}
\usepackage{pslatex}
\usepackage{algorithm2e}
\usepackage[bottom]{footmisc}
\usepackage{natbib}

\usepackage{graphicx}%
\graphicspath{{./figures/}}
\usepackage{amsmath,amssymb,amsfonts,amsthm}%
\usepackage{xurl}

\usepackage{xcolor}%

\usepackage{tikz,lipsum,lmodern}
\usetikzlibrary{matrix,arrows.meta, positioning, fit, backgrounds}
\usepackage{adjustbox}

\usepackage{enumerate}
\usepackage{caption}
\usepackage{subcaption}
\usepackage[most]{tcolorbox}
\newcommand{\ket}[1]{\ensuremath{\left|#1\right\rangle}} 
\newcommand{\bra}[1]{\ensuremath{\left\langle#1\right|}} 

\usepackage{textcomp}
\usepackage{authblk}
\usepackage{orcidlink}
\date{}
\begin{document}

\title{Quantum RNNs and LSTMs Through Entangling and Disentangling Power of Unitary Transformations}

\author{Ammar Daskin\orcidlink{0000-0002-1497-5031}\thanks{adaskin25@gmail.com}}
\affil{Department of Computer Engineering, Istanbul Medeniyet University, Uskudar, Istanbul, Turkiye}

\maketitle

\abstract{In this paper, we present a framework for modeling quantum recurrent neural networks (RNNs) and their enhanced version, long short-term memory (LSTM) networks using the core ideas presented by Linden et al. (2009), where the entangling and disentangling power of unitary transformations is investigated. In particular, we interpret entangling and disentangling power as information retention and forgetting mechanisms in LSTMs. 
Thus, entanglement emerges as a key component of the optimization (training) process. We believe that, by leveraging prior knowledge of the entangling power of unitaries, the proposed quantum-classical framework can guide the design of better-parameterized quantum circuits for various real-world applications.}


\section{\uppercase{Introduction}}  
Quantum information is carried by qubits, implemented by the basic any two‑level quantum systems. The pure state for such a system can be written in two-dimensional Hilbert space as \(\ket{\psi}=\alpha\ket{0}+\beta\ket{1}\) with complex amplitudes \(\alpha,\beta\) satisfying the normalization condition $|\alpha|^2+|\beta|^2=1$ and $\ket{0}$ and $\ket{1}$ being the Dirac notation for standard two dimensional basis vectors (note that when the state is mixed, the state description comes with a summation formula in such basis state). The state of an \(n\)-qubit register lies in a \(2^n\)-dimensional Hilbert vector space. Any physical transformations in this space preserves the unity norm and can be described by unitary matrices. While this exponential growth of the state space underlies the difficulty of simulating quantum systems on classical computers,  it provides a representational power for quantum circuits in the hope that we can solve many problems more efficiently \citep{nielsen2010quantum}.

A central property that distinguishes quantum systems from classical ones and imposes the simulation difficulty is the entanglement which describes the nonclassical correlation that can exist between subsystems. In mathematical language, the entangled states cannot be described by using only local representation through Kronecker tensor product, $\otimes$. For instance, $\alpha \ket{00}+\beta \ket{11}$ cannot be explained as a tensor product of the individual qubit states. However, when we measure (i.e., attempt to obtain a classical value from) one qubit in any of $\ket{0}$ or $\ket{1}$ state, then we can be sure what the state of the other qubit is.
For a bipartite pure state \(\ket{\psi_{AB}}\) described in Hilbert space $A\otimes B$, the amount of the entanglement is formulated by the reduced density matrix:
\begin{equation}
    \rho_A=\mathrm{Tr}_B(\ket{\psi_{AB}}\bra{\psi_{AB}}).
\end{equation}
This can be used to describe the probabilities of the measurement outcomes of the isolated subsystems. The entanglement can be also identified by the von Neumann entropy:
\begin{equation}
\label{entropy}
    S(\rho_A)=-\mathrm{Tr}(\rho_A\log_2\rho_A).
\end{equation} 
If \(S(\rho_A)>0\) the subsystems are entangled; if \(S(\rho_A)=0\) the state is separable. The Schmidt decomposition provides an explicit form \(\ket{\psi_{AB}}=\sum_i\sqrt{\lambda_i}\ket{i_A}\otimes\ket{i_B}\), where the \(\lambda_i\) (the eigenvalues of \(\rho_A\)) determine the entanglement spectrum and directly relate to the von Neumann entropy \citep{nielsen2010quantum}.

Entanglement also has practical consequences for quantum algorithms and circuit design. Highly entangling unitaries can generate states that are classically hard to simulate \citep{bravyi2018quantum} and can increase expressivity in variational circuits, while limited entanglement often constrains representational capacity \citep{schuld2021effect,levine2019quantum,du2020expressive}. 
Conversely, the ability of a unitary to entangle or disentangle subsystems-formalized as entangling and disentangling power \citep{linden2009entangling}-affects how information is stored, propagated, and erased across registers \citep{nielsen2010quantum}. This makes entanglement a natural quantity to consider when designing quantum recurrent architectures and memory models.

\begin{figure*}[!h]
  \vspace{-0.2cm}
  \centering
    \includegraphics[width=0.9\textwidth]{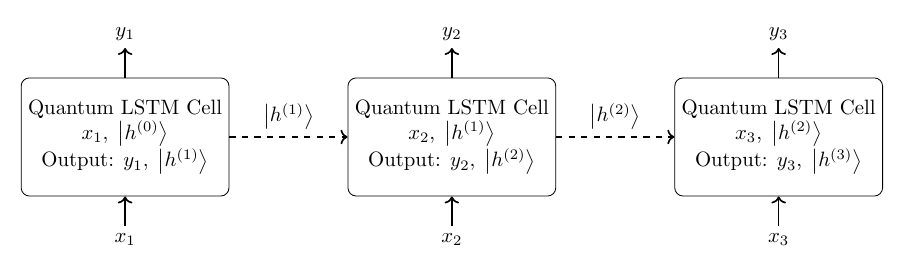}
    \caption{Generic Structure of Quantum LSTM.}
    \label{fig:quantumLSTM}
\end{figure*}

For a unitary \( U \) that transforms an input state \( \ket{\Psi_{\rm in}} \) to an output state \( \ket{\Psi_{\rm out}} \) through \( \ket{\Psi_{\rm out}} = U\ket{\Psi_{\rm in}} \), its entangling and disentangling powers, respectively, \( E^\uparrow(U) \) and \( E^\downarrow(U) \), are defined as the maximum increase and decrease in entanglement between two subsystems after applying \( U \) \citep{linden2009entangling}:
\begin{equation}
\begin{split}
    E^\uparrow(U) &= \text{sup}_{\ket{\Psi_{\text{in}}}} \left( E(\Psi_{\text{out}}) - E(\Psi_{\text{in}}) \right),\\
    E^\downarrow(U) &= \text{sup}_{\ket{\Psi_{\text{in}}}} \left( E(\Psi_{\text{in}}) - E(\Psi_{\text{out}}) \right).
\end{split}
\end{equation}
\citet{linden2009entangling} have shown that for bipartite systems, although the entangling and disentangling powers are the same for two-qubit unitaries, they differ in general for unitaries of larger dimensions. As an example, they have demonstrated a qubit-qutrit unitary gate \( U_{2\times3} \) that can create 2 ebits of entanglement; however, the system cannot be fully disentangled using a similar unitary. Notably, for the inverse transformation \( U^\dagger \), the relation \( E^\downarrow(U) = E^\uparrow(U^\dagger) \) holds since \( U^\dagger \) reverses the transformation effect of \( U \).  

This entangling power has been analytically investigated for bipartite permutation matrices \citep{chen2016entangling} and two-qubit mixed states \citep{guan2014entangling}. It has known upper bounds \citep{bravyi2007upper,das2020entanglement} and can be maximized using perfect entangling gates \citep{musz2013unitary,yu2010optimal} or by incorporating ancilla qubits \citep{linden2009entangling}. Additionally, it is relevant to communication capacity \citep{harrow2005time,bauml2018fundamental} and has an influence over quantum circuit complexity \citep{eisert2021entangling}.

The entangling and disentangling power of unitary transformations provide a framework for constructing a memory model in which entanglement can be explicitly quantified using previously analyzed unitaries. Consequently, a quantum machine learning model designed around the entangling power of unitaries can be studied more easily when we try to understand its expressivity and demonstrate its fundamental quantum advantages over classical counterparts.

The recurrent neural network (RNN) or long short-term memory (LSTM) cell models require a dynamic memory to retain or forget previous information. Previous works such as \citep{chen2022quantum} or its quantum kernel-based version \citep{hsu2024quantum} use variational quantum circuits (VQCs) as subcomponents within a classical LSTM structure. The gates (forget, input, output) are replaced by VQCs, but the overall architecture remains mostly classical due to the use of classical memory. There are also several applications and comparison studies of these models with various variational circuit choices \citep{khan2024quantum,saha2025advancing,choppara2025leveraging,tripathi2025quantum,mahmood2024comparative,padha2024qclr}. \citet{zhou2024implementation} discusses different design choices and provides implementation guidelines. \citet{ceschini2021design} presents a fully quantum LSTM model, where the layers of a standard LSTM are cascaded using ancilla registers and quantum multiplication and addition operations.

In this paper, we discuss how quantum RNNs and LSTMs can be modeled using the core ideas presented by \citet{linden2009entangling}, where the entangling and disentangling power can be directly related to information retention and forgetting, respectively. In particular, we consider the environment (ancilla) qubits as quantum memory representing the temporal hidden state of the model. Therefore, the quantum register acts as memory, storing the history of the hidden state.

The optimization process in this model also involves optimizing how much entanglement is retained in the circuit. Although similar approaches can be observed in the previously discussed papers, we believe our work may help advance understanding how entanglement plays a role in carrying and propagating information in quantum machine learning models, especially LSTMs. Therefore, it can serve as a guide for determining a parameterized circuit choice in different applications, making entanglement a direct component of the learning process.

In the following sections, after explaining the general framework of the LSTM and its single LSTM cell in detail, we present numerical simulations on noisy random sine data to gauge the expressive power of the model. In addition, we present simulation results for weather data, demonstrating the model’s applicability to real-world scenarios.

\begin{figure*}[!h]
  \vspace{-0.2cm}
  \centering
    \includegraphics[width=0.9\textwidth]{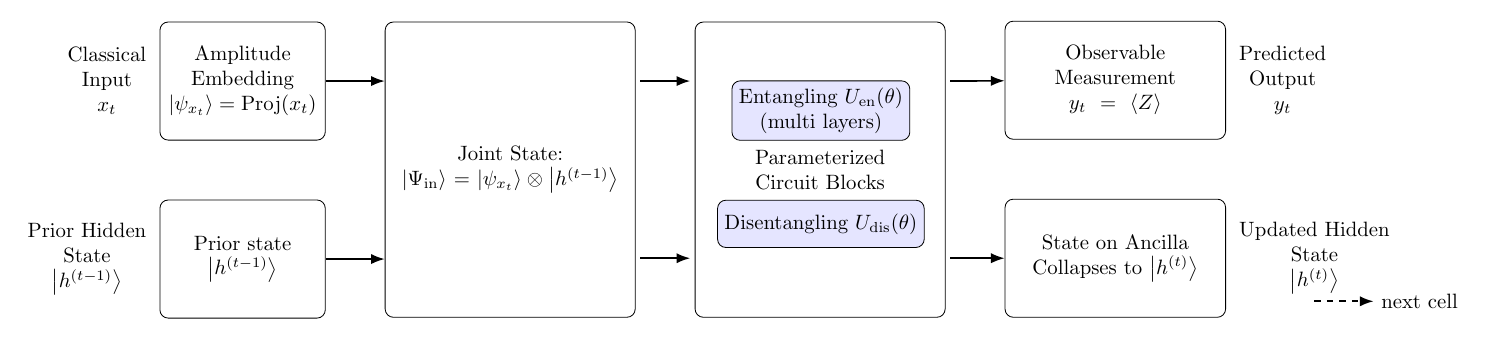}
    \caption{A Single Quantum LSTM Cell.}
    \label{fig:quantumLSTMcell}
\end{figure*}

\section{\uppercase{The Hybrid General Framework}}
In a practical time-series task (such as forecasting), the model processes an ordered sequence of inputs: \( \{x_1, x_2, \dots, x_T\}\), where each \( x_t \) is a scalar (or a feature vector) at time \( t \). Our classical-quantum hybrid framework is based on using a quantum register for the input data \( x_t \) and an ancilla register for the hidden state \( h_t \).  

One of the core principles in RNNs and LSTMs is the ability to sequentially process the given data, i.e., propagate information across time steps, which is referred to as hidden information. In our design, an ancilla register is used as a quantum memory to hold this hidden historical knowledge through its remaining entanglement with the system register from previous steps.

This general framework is depicted in Fig.~\ref{fig:quantumLSTM}. Each cell in this LSTM model represents an iteration, i.e., an application of a parameterized circuit to different inputs. The explicit details of an LSTM cell are given in Fig.~\ref{fig:quantumLSTMcell}. A high-level description for time step $t$ can be enumerated as follows:
\begin{enumerate}
    \item 
\textbf{Inputs}: The inputs to a cell are the classical data input \( x_t \) and the quantum hidden state \( \ket{h^{(t-1)}} \), which propagates from the previous time step \( t-1 \).  
    \begin{enumerate}
        \item Since the classical input at each time step is a scalar value, it is projected onto a high-dimensional quantum state using an affine transformation \citep{bergholm2018pennylane,schuld2021machine}. In particular, a classical scalar \( x_t \in \mathbb{R} \) is mapped by a linear layer to produce a vector \( \mathbf{a_t} = W_{x_t}\, x_t \).  
        This vector is then normalized to form a quantum state in the system subspace:
        \begin{equation}
        \ket{\psi_x} = \frac{1}{\|\mathbf{a}_t\|}\mathbf{a}_t \quad \in \mathcal{H}_{\text{sys}}, 
        \end{equation}
        where \( \dim(\mathcal{H}_{\text{sys}}) = 2^{n_{\text{sys}}} \) represents the Hilbert space for the system register. Note that in our simulations, we have used \( n_{\text{sys}} = 2 \).  
        \item The hidden state is represented as a quantum state in the ancilla subspace and is typically initialized to a standard basis state. For example, we have used the following initial setting in our simulations:  
        \begin{equation}
        \ket{h^{(0)}} = \ket{0}^{\otimes n_{\text{ancilla}}}=  \begin{pmatrix} 1 \\ 0 \\ \vdots \\ 0 \end{pmatrix} .
        \end{equation}
    \end{enumerate}

\item    
\textbf{The input quantum state}: The whole joint state is given by:
    \begin{equation}
      \ket{\Psi_{\rm in}^{(t)}} = \ket{\psi_{x_t}} \otimes \ket{h^{(t-1)}} \quad \in \mathcal{H}_{\text{sys}} \otimes \mathcal{H}_{\text{ancilla}}. 
    \end{equation}
    Here, \( \mathcal{H}_{\text{ancilla}} = 2^{n_{\text{ancilla}}} \) represents the Hilbert space for the ancilla register, and in our simulations, we have used \( n_{\text{ancilla}} = 2 \). (In our simulations, the system and ancilla have the same sizes; however, this is not a requirement.) Note that this joint state is embedded into the quantum circuit via amplitude encoding techniques.  
\item 
\textbf{Parameterized circuit}: We then apply two similarly structured quantum circuits, \( U_{\mathrm{en}} \) and \( U_{\mathrm{dis}} \), with different parameters to emulate entangling and disentangling dynamics:
    \begin{equation}
    \ket{\Psi_{\rm out}^{(t)}} = U_{\mathrm{dis}}\, U_{\mathrm{en}}\, \ket{\Psi_{\rm in}^{(t)}}.
    \end{equation}
    In the simulations, we have used a basic two-layer entangling circuit. One can increase the number of parameters in the model by choosing more complex designs. Note that, since the entangling and disentangling powers are different, we use similarly structured circuits with different trainable parameters in the hope that, by the end of the learning process, entanglement retention will be optimized.
\item     
\textbf{Final state and output extraction}: After the evolution by the parameterized circuit, the full state can be written in terms of its subspaces as:
    \begin{equation}
    \ket{\Psi_{\rm out}}^{(t)} = \sum_{i=0}^{2^{n_{\rm sys}}-1} \ket{i}_{\rm sys} \otimes \ket{\phi_i}_{\rm ancilla},   
    \end{equation}
    where \( \ket{i} \) represents the \( i \)th basis in the standard basis, and each \( \ket{\phi_i} \) is an (unnormalized) state in the hidden (ancilla) Hilbert space. In quantum theory, after measurement, the quantum state becomes a mixed state that can be described by the reduced density matrix. In numerical simulations, one can obtain the reduced density matrix for the ancilla by tracing out the system qubits \citep{nielsen2010quantum}:
    \begin{equation}
    \rho_{\text{ancilla}} = \mathrm{Tr}_{\text{sys}}\left(\ket{\Psi_{\text{out}}^{(t)}}\bra{\Psi_{\text{out}}^{(t)}}\right).
    \end{equation}

    The state \( \ket{\phi_i} \) on the ancilla determines the probability \( p(i) = \|\ket{\phi_i}\|^2 \) associated with the system outcome \( \ket{i}_{\rm sys} \) on the system register.  
    Therefore, if we choose the \( i^* \) corresponding to the highest probability, then upon measurement of the system register, the state on the ancilla collapses to \( \ket{\phi_{i^*}} \):
    \begin{equation}
    \ket{\Psi_{\rm out}^{(t)}} \quad \overset{\text{Measure}}{\longrightarrow} \quad \ket{i^*}_{\rm sys} \otimes \ket{h^{(t)}},
    \end{equation}
    Thus, the hidden state is updated with the normalized ancilla component:
    \begin{equation}
    \label{eq_hidden}
    \ket{h^{(t)}} = \frac{\ket{\phi_{i^*}}}{\|\ket{\phi_{i^*}}\|}.
    \end{equation}

Alternatively, in the simulations, we have also utilized the normalized diagonal of the density matrix:
  \begin{equation}
        \ket{h^{(t)}}= \operatorname{diag}\left(\rho_{\text{ancilla}}\right).
  \end{equation}
  While \( \ket{\phi_{i^*}} \) is composed of amplitudes, this vector contains the probabilities \( [p_0, p_1, \dots, p_{2^{n_{\text{ancilla}}}-1}] \), where \( p_j = \bra{j}\rho_{\text{ancilla}}\ket{j} \).  
  On a quantum machine, when the size of the ancilla register is small (as in our simulations, where it consists of only a few qubits), one can approximate the full tomography of the state by using a few repetitions of the circuit. However, for larger ancilla sizes, a more computationally efficient choice would be to measure individual qubits and combine their results to obtain an approximate representation of the previous hidden state.

  Note that transitioning from amplitudes to probabilities, and then to individual probabilities (or binary outcomes), inherently leads to information loss. The appropriate choice can be determined based on the application, as some applications may tolerate such information loss (e.g., data points that are less likely to depend on previously observed values).

Alongside updating the hidden state, the circuit also computes an expectation value of an observable on one of the system qubits. The chosen qubit (or qubits if \( y_t \) is a vector) can be selected from among the system qubits. Then, the predicted output is:
\begin{equation}
    y_t = \langle \Psi_{\text{out}}^{(t)}| Z_{m} |\Psi_{\text{out}}^{(t)} \rangle,
\end{equation}
where \( Z_m \) denotes a measurement operator, e.g., the Pauli-\( Z \) operator on a chosen qubit.

\item 
\textbf{Optimization (training)}: The sequence of outputs \( \{y_t\} \) is collected to form the output of the cell for that input sequence. A classical loss function (we use mean squared error (MSE) in numerical simulations) is computed between the predicted outputs and the target values. The network is trained end-to-end via backpropagation, even though the collapse operation simulating an actual projective measurement is non-differentiable. However, it can be estimated using the parameter-shift rule \citep{schuld2021machine}.
\end{enumerate}

As explained in the above steps and as also shown in Fig.~\ref{fig:quantumLSTM}, propagating information to the next cell is done through a collapsed quantum state or by obtaining a partial tomography. Collapsing a quantum state is achieved by measuring the system register, which can involve sequentially aligned individual qubit measurements or a full quantum register measurement.  
Therefore, it is important to point out that the required number of qubits is directly related to the circuit output (which, for time-series tasks, is generally a scalar) and the number of parameters used to train the model. If the quantum architecture allows such mid-level quantum circuit measurements, then this model can be implemented directly on such an architecture, as it would require only a limited number of qubits.

\section{\uppercase{Analysis of the Model to Show Why It Works}}  
To establish the formal connection between unitary action, entanglement change, and ancilla memory behavior that underpins the quantum LSTM model, in this section we provide an analysis based on subsystem entropies of the subsystems.
We aim to formalize how unitary-induced entanglement-changes govern memory retention and forgetting in the quantum LSTM.

As given in Eq.\eqref{entropy}, von Neuman entropy can be used to understand entanglement. In our model, let us assume that \(\mathcal H_{\rm sys}\) and \(\mathcal H_{\rm anc}\) denote the system and ancilla Hilbert spaces with dimensions \(2^{n_{\rm sys}}\) and \(2^{n_{\rm anc}}\). Therefore, we can denote the input state as joint pure state at time \(t\) by \(\ket{\Psi_{\rm in}^{(t)}}\in\mathcal H_{\rm sys}\otimes\mathcal H_{\rm anc}\). 
Since the parameterized evolution inside a cell is a unitary \(U=U_{\mathrm{dis}}U_{\mathrm{en}}\); for any pure input \(\ket{\Psi_{\rm in}}\), the output \(\ket{\Psi_{\rm out}}=U\ket{\Psi_{\rm in}}\) is pure and normalized. 

For time \(t\), we can write the reduced density matrices for the joint output state \(\ket{\Psi_{\rm out}^{(t)}}=U\ket{\Psi_{\rm in}^{(t)}}\) as:
\begin{equation}
\begin{split}
&\rho_{\rm sys}=\mathrm{Tr}_{\rm anc}(\ket{\Psi_{\rm out}^{(t)}}
\bra{\Psi_{\rm out}^{(t)}
}),\\
&\rho_{\rm anc}=\mathrm{Tr}_{\rm sys}(\ket{\Psi_{\rm out}^{(t)}}\bra{\Psi_{\rm out}^{(t)}}).
\end{split}
\end{equation}
It is known that entropy of the subsystems equals across bipartition for pure joint states \citep{nielsen2010quantum}. 
Therefore, when the input or output \(\ket{\Psi}\) is pure on \(\mathcal H_{\rm sys}\otimes\mathcal H_{\rm anc}\), we have \(S(\rho_{\rm sys})=S(\rho_{\rm anc})\).

 The bound on the change of entropy can be understood through the entangling and disentangling power of unitaries \citep{linden2009entangling}. Consider applying unitary $U$ to an arbitrary pure input state \(\ket{\Psi_{\rm in}}\), we can define the entropy change on the ancilla as \(\Delta S_{\rm anc}=S(\rho_{\rm anc}^{\rm out})-S(\rho_{\rm anc}^{\rm in})\).  By definition, we can write the following:
 \begin{equation}
E^\uparrow(U)=\text{sup}_{\ket{\Psi_{\rm in}}}\big(S(\rho^{\rm out})-S(\rho^{\rm in})\big),
 \end{equation}
where the entropy is taken on the chosen subsystem. 
For any particular  \(\ket{\Psi_{\rm in}}\) the realized change \(\Delta S_{\rm anc}\) cannot exceed the supremum, hence,
 \begin{equation}
\Delta S_{\rm anc}\le E^\uparrow(U). 
 \end{equation}
 Applying the same definition to \(U^\dagger\) and using 
 \(E^\downarrow(U)=E^\uparrow(U^\dagger)\) yields the lower bound
 \begin{equation}
      \Delta S_{\rm anc}\ge -E^\downarrow(U). 
 \end{equation}
 From these two, we can see that the entangling power \(E^\uparrow(U)\) and disentangling power \(E^\downarrow(U)\) bound the possible positive and negative values of \(\Delta S_{\rm anc}\): 
\begin{equation}
    -\;E^\downarrow(U)\ \le\ \Delta S_{\rm anc}\ \le\ E^\uparrow(U).
\end{equation}
These bounds show that $E^\uparrow(U)$ quantifies the maximum possible memory creation (entanglement increase), while $E^\downarrow(U)$ quantifies the maximum possible memory erasure (entanglement decrease).

\textbf{Measurement update and ancilla memory (collapse and averaged map):}
Expressing the joint output state in a Schmidt-like decomposition over system basis, we can write:
\begin{equation}
\ket{\Psi_{\rm out}}=\sum_i \ket{i}_{\rm sys}\otimes\ket{\phi_i}_{\rm anc}.
\end{equation}
From this point, measuring the system and obtaining outcome \(i\) projects the joint state to \(\ket{i}\otimes\ket{\phi_i}\). Normalizing yields the conditional ancilla state \(\ket{\phi_i}/\|\ket{\phi_i}\|\) which gives Eq.\eqref{eq_hidden}. As explained in the previous section, we can also do the nonselective post‑measurement on ancilla through the density matrix
\begin{equation}
    \rho_{\rm anc}^{\rm out}=\sum_i \ket{\phi_i}\bra{\phi_i}=\sum_i \big({}_{\rm sys}\!\bra{i}U\big)\,\rho_{\rm anc}^{\rm in}\,\big(U^\dagger\ket{i}_{\rm sys}\big),
\end{equation}
which is the Kraus form with \(K_i={}_{\rm sys}\!\bra{i}U\). The map is the completely positive trace preserving (CPTP) map because \(\sum_i K_i^\dagger K_i=I_{\rm anc}\) follows from unitarity of \(U\) and completeness of the measurement basis. Therefore the ancilla evolves according to the  following map:
\begin{equation}
    \rho_{\rm anc}^{\rm out}=\sum_i K_i\,\rho_{\rm anc}^{\rm in}\,K_i^\dagger.
\end{equation}

In order to understand pre- and post-measurement hidden states, we can use the fidelity (the closeness of two system) which is described for the quantum systems $\rho$ and $\sigma$ as:
 \begin{equation}
   F(\rho,\sigma)=\big(\mathrm{Tr}\sqrt{\sqrt\rho\sigma\sqrt\rho}\big)^2.  
 \end{equation}
Fidelity quantifies state similarity, with $F=1$ for identical states and $F=0$ for orthogonal states. 
Thus, by adjusting $U$'s parameters during training, the model learns to modulate entanglement—and consequently, fidelity with previous hidden states—to implement the desired balance of memory retention and forgetting for the specific task.

On the other hand, entanglement-increase between system and ancilla implies that the reduced state on the ancilla becomes more mixed (higher entropy) for many inputs, which typically reduces fidelity with the original ancilla state. 
The fidelity is known to be monotone with trace distance, an increase in subsystem entropy (bounded by \(E^\uparrow(U)\)) implies a lower bound on the trace distance and hence an upper bound on fidelity \citep{miszczak2008sub}. 
  
Therefore, for the ancilla hidden state \(\rho_{\rm anc}^{(t-1)}\) at time \(t-1\); applying \(U\) and performing the measurement/selection described above, the fidelity between the pre‑ and post‑hidden states (averaged over measurement outcomes) is bounded below by a function of the entanglement change: Large positive entangling power generally reduces average fidelity between \(\rho_{\rm anc}^{(t-1)}\) and \(\rho_{\rm anc}^{(t)}\).
Consequently, we can expect unitaries with large entangling power can reduce memory fidelity on average, while unitaries with small entangling power tend to preserve ancilla memory.
Furthermore, by utilizing entanglement as a controllable memory resource and parameterizing \(U\)  during training, the model can tune the expected entropy change \(\mathbb E[\Delta S_{\rm anc}]\) to match the memory retention and forgetting requirements in the learning task.

\section{\uppercase{Numerical Simulations}}

\begin{figure}[!h]
  \vspace{-0.2cm}
  \centering
    \includegraphics[width=\linewidth]{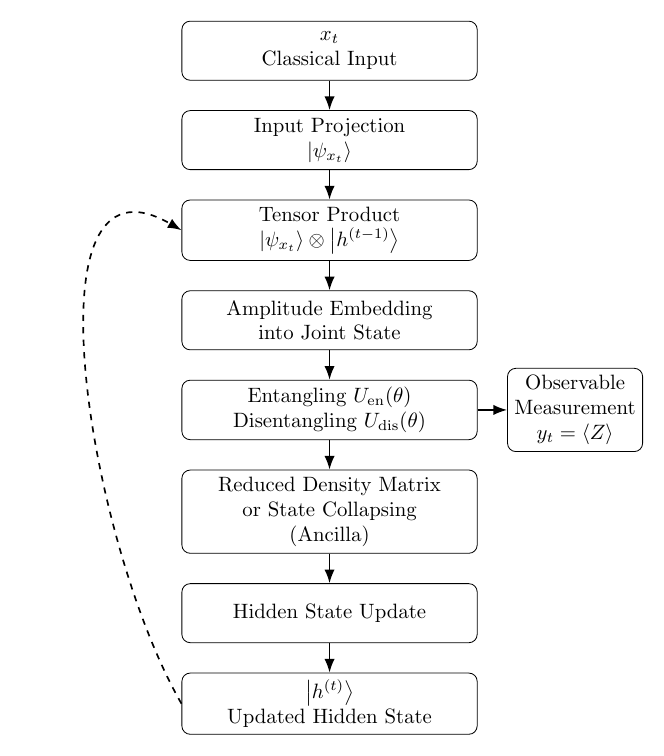}
    \caption{Numerical Steps Used to Simulate Quantum LSTM.}
    \label{fig:FlowChart}
\end{figure}

Here, note that the simulation code is available for download, and the results can be regenerated using the link provided in the data availability section. In the code, the flowchart given in Fig.~\ref{fig:FlowChart} is implemented in two different ways: separately for the reduced density matrix formalism and the collapsed state formalism.  

The simulation code uses the Pennylane \citep{bergholm2018pennylane} Python library for generating quantum circuits and its machine learning libraries, along with PyTorch for training the model. For the entangling and disentangling gates, we have used the Pennylane quantum library to generate a basic entangling gate, with a similar structure used for the disentangling gate. Both of these operators use separate trainable parameters.  

To study the expressive power of the model, we have used a noisy sine function. To demonstrate its applicability to real-world problems, we have used one year of weather data from Ontario, Canada.

\begin{figure*}[!h]
  \vspace{-0.2cm}
  \centering
\begin{subfigure}{0.45\textwidth}
    \centering
    \includegraphics[width=1\linewidth]{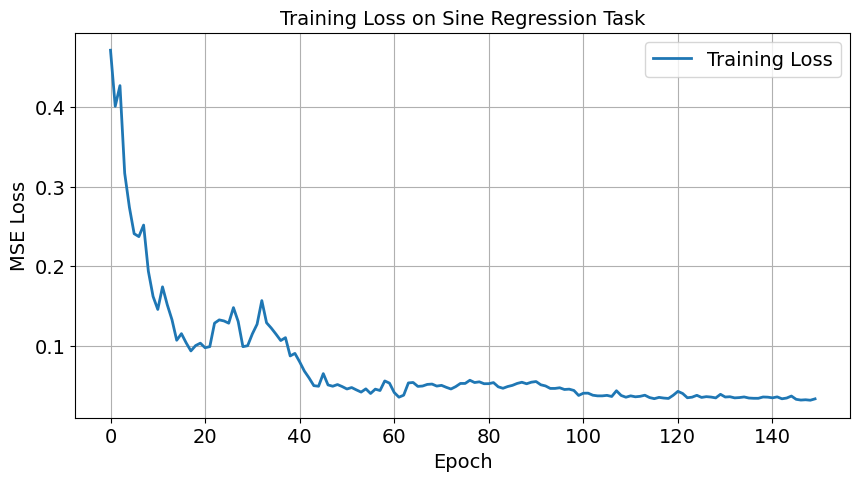}
    \caption{Loss when collapsed state is used for hidden state}    
\end{subfigure}\hfill 
\begin{subfigure}{0.45\textwidth}
    \centering
    \includegraphics[width=1\linewidth]{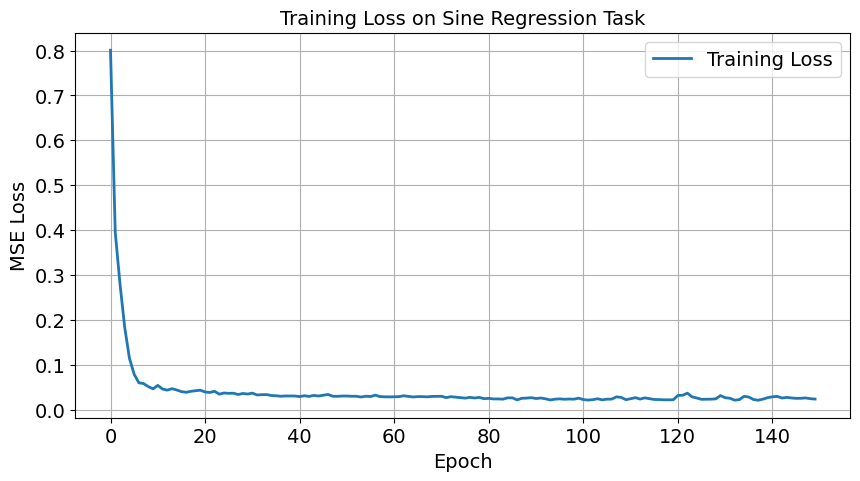}
    \caption{Loss when probabilities are used for hidden state}    
\end{subfigure}\\
\begin{subfigure}{0.45\textwidth}
    \centering
    \includegraphics[width=1\linewidth]{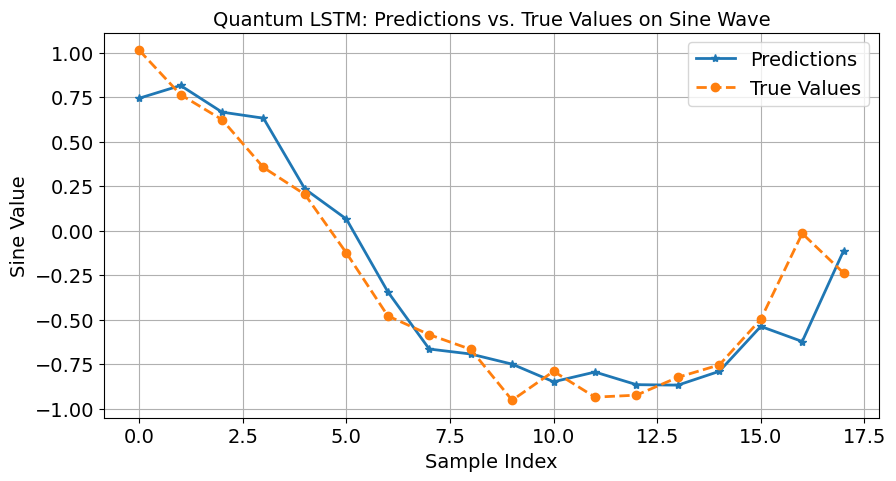}
    \caption{Predictions vs true values for the test cases when collapsed state is used for hidden state}    
\end{subfigure}\hfill
\begin{subfigure}{0.45\textwidth}
    \centering
    \includegraphics[width=1\linewidth]{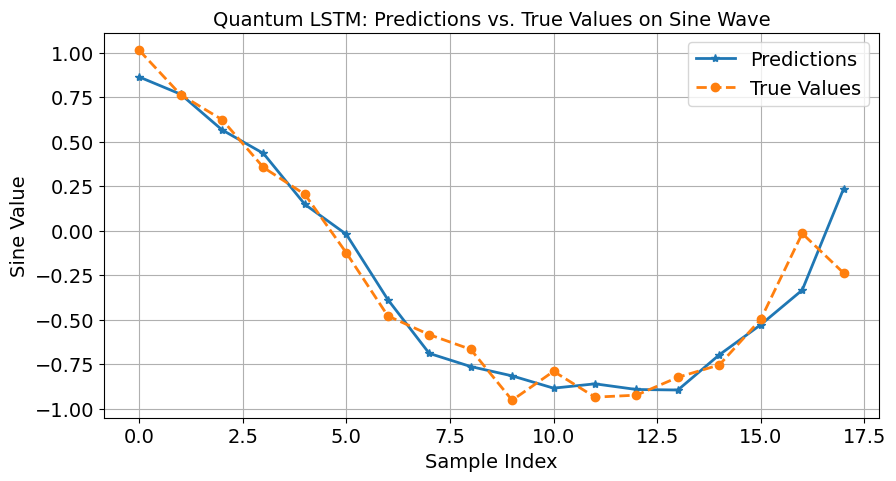}
    \caption{Predictions vs true values for the test cases when probabilities are used for hidden state}    
\end{subfigure}
    \caption{The loss and the predictions vs true values in the test cases of the noisy sine function for the values in range $[0, 8\pi]$. }
    \label{fig:sinfunction}
\end{figure*}

\begin{figure*}[!h]
  \vspace{-0.2cm}
  \centering
\begin{subfigure}{0.45\textwidth}
    \centering
    \includegraphics[width=1\linewidth]{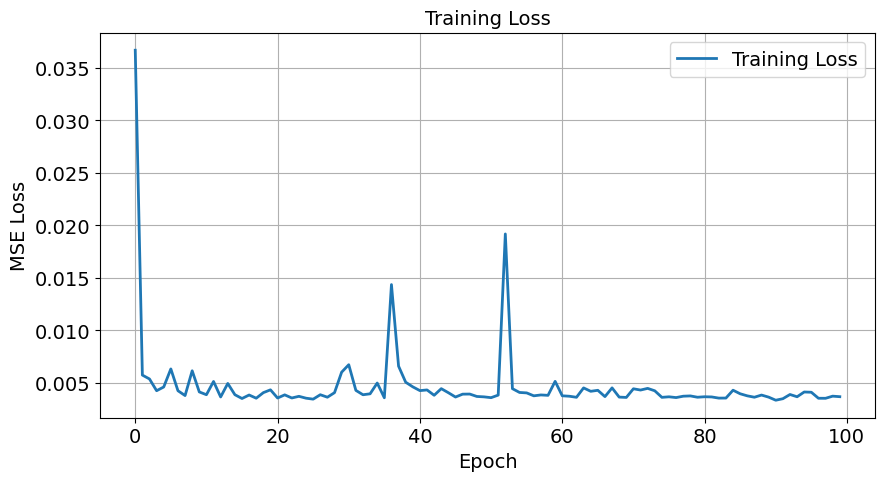}
    \caption{Loss when collapsed state is used for hidden state}    
\end{subfigure}\hfill 
\begin{subfigure}{0.45\textwidth}
    \centering
    \includegraphics[width=1\linewidth]{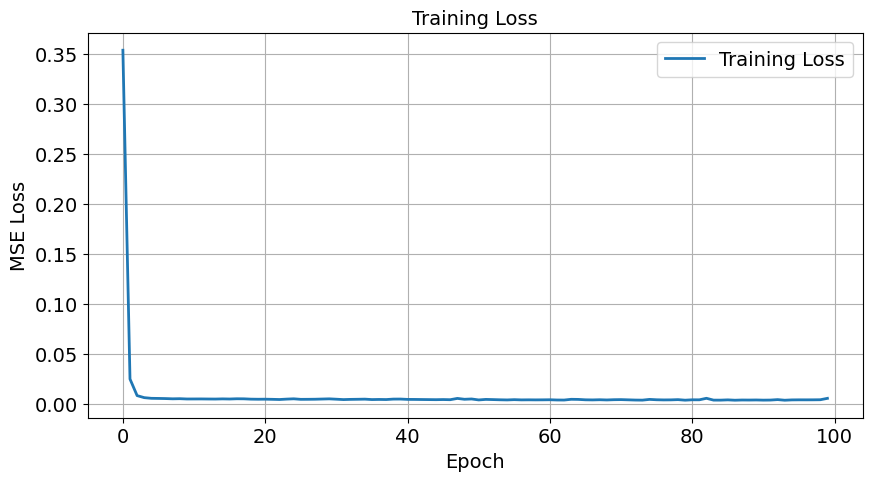}
    \caption{Loss when probabilities are used for hidden state}    
\end{subfigure}\\
\begin{subfigure}{0.45\textwidth}
    \centering
    \includegraphics[width=1\linewidth]{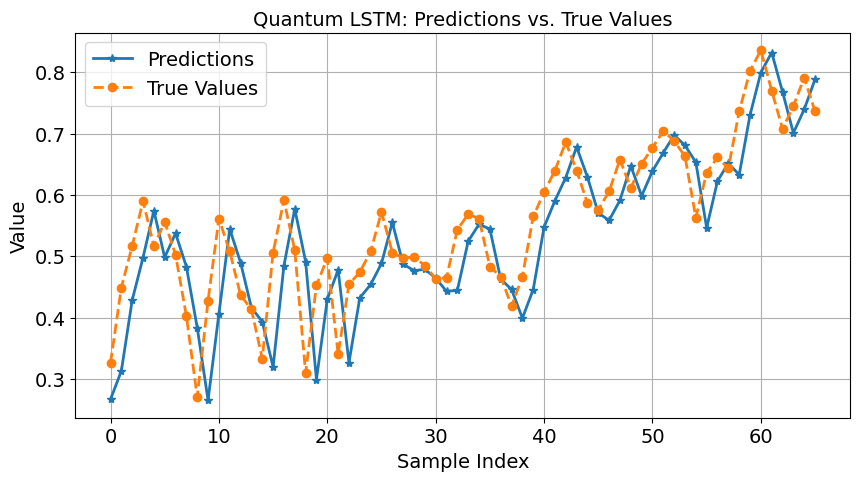}
    \caption{Predictions vs true values for the test cases when collapsed state is used for hidden state}    
\end{subfigure}\hfill
\begin{subfigure}{0.45\textwidth}
    \centering
    \includegraphics[width=1\linewidth]{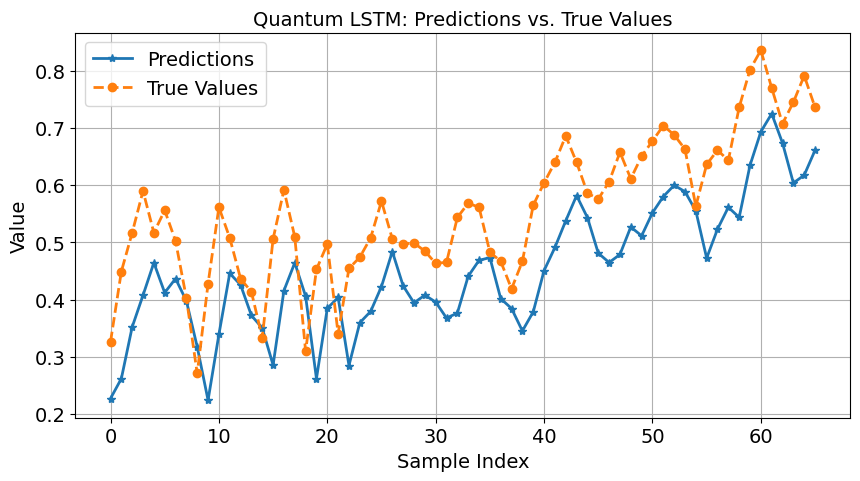}
    \caption{Predictions vs true values for the test cases when probabilities are used for hidden state}    
\end{subfigure}
    \caption{The loss and the predictions vs true values in the test cases of the weather data for Ontario, Canada. }
    \label{fig:weather}
\end{figure*}

\begin{itemize}
    \item The noisy sine data is generated by first creating 100 values in the range \( [0, 8\pi] \), computing their sine, and then adding different random noise between \( [-0.1, 0.1] \) to each function value. As in classical training schemes, the data is divided into two partitions: \( 80\% \) for training and \( 20\% \) for testing. We use a batch size of 5 and 0.01 for the learning rate. Following random sampling, \( 80\% \) of the data is used for training, and the remaining \( 20\% \) is used for testing. For both the reduced density matrix and the collapsed state computations, Fig.~\ref{fig:sinfunction} shows the loss values during training, as well as the predicted and true values for the test portion of the data (the \( 20\% \) separated part).  

    \item We have also taken 365 days of weather data between 10 May 2024 and 10 May 2025 for Ontario, Canada \citep{meteostat_website,meteostat_python}. With settings similar to those used for the sine function, we run the simulations. The results are reported in Fig.~\ref{fig:weather}.  
\end{itemize}

As seen from both figures, the model is mostly able to predict the expected values. We also observe that when using the collapsed state, the loss exhibits occasional sharp increases. This may be a good indication that the model has the ability to jump over local minima when it gets stuck. As mentioned before, for more complex datasets; accuracy can be improved by using deeper parameterized circuits or increasing the number of qubits. Both of these enhance the expressive power of quantum machine learning models but make training more difficult \citep{mcclean2018barren,larocca2025barren}.

\section{\uppercase{Conclusion}}
In this paper, we have described a quantum LSTM model in which the entanglement between system and ancilla registers directly encodes temporal memory by mimicking the entangling and disentangling power difference of unitary transformations. In existing quantum LSTM models, entanglement primarily enhances nonlinear expressivity rather than explicitly storing temporal information. In our model, however, entanglement is explicitly considered as part of the optimization process. Consequently, our framework can leverage analytical insights of the entangling power of unitary transformations to design application-specific parameterized circuits and quantum machine learning models.

\section{\uppercase{Data availability}}
The simulation code used to generate all figures presented in this paper is publicly available
on: \url{https://github.com/adaskin/quantum-lstm}

\section{\uppercase{Funding}}
This project is not funded by any funding agency.

\section{\uppercase{Acknowledgment}}
When writing simulation code, we acknowledge that we have benefited suggestions from various AI tools mainly: Microsoft Copilot \citep{Copilot2025} and DeepSeek-R1/V3 \citep{DeepSeek2025} . And the paper has been proofread, without changing the structures of the paragraphs and sentences by using these tools.

\newpage
\bibliographystyle{apalike}
{\small
\bibliography{main}}

\end{document}